\documentclass[prd,aps,nofootinbib,preprintnumbers]{revtex4}
\usepackage{amssymb}
\usepackage{bm}
\usepackage{enumerate}
\usepackage{amssymb}
\usepackage{amsmath}
\usepackage{feynmf}
\usepackage{slashed}
\usepackage{wrapfig}
\usepackage{subfig}
\usepackage{graphicx}
\usepackage{filecontents}

\newcommand\be{\begin{equation}}
\newcommand\ee{\end{equation}}

\begin{document}

\title 
{\Large \bf Non-thermal Dark Matter from Modified Early Matter Domination} 

\author{Rouzbeh Allahverdi$^{1}$}
\author{Jacek K. Osi{\'n}ski$^{1}$}

\affiliation{$^{1}$~Department of Physics and Astronomy, University of New Mexico, Albuquerque, NM 87131, USA}

%%%%%%%%%%%%%%%%%%%%%%%%%%%%%%%%%%%%%%%%%%%%%%%%%%%%%%%%%%%
\begin{abstract}
Thermal freeze-out or freeze-in during a period of early matter domination can give rise to the correct dark matter abundance for $\langle \sigma_{\rm ann} v \rangle_{\rm f} < 3 \times 10^{-26}$ cm$^3$ s$^{-1}$. In the standard scenario, a single field that behaves like matter drives the early matter dominated era. However, in realistic models, this epoch may involve more than one field. In this paper, we study the effect of such a modification on the production of dark matter during early matter domination. We show that even a subdominant second field that decays much faster than the dominant one can considerably enhance the temperature of the universe during an early matter-dominated phase. This in turn affects dark matter production via freeze-out/in and opens up the allowed parameter space toward significantly larger dark matter masses. As a result, one can comfortably obtain the correct relic abundance for PeV-scale dark matter for reheating temperatures at or below 10 GeV.      
\end{abstract}
\maketitle

%%%%%%%%%%%%%%%%%%%%%%%%%%%%%%%%%%%%%%%%%%%%%%%%%%%%%%%%%%%%
\section{Introduction}

There are various lines of evidence that most of the matter in the universe is dark~\cite{BHS}. However, the identity of dark matter (DM) remains as a major problem at the interface of cosmology and particle physics. Weakly interacting massive particles (WIMPs) are promising candidates for DM and have been the main focus of direct, indirect, and collider searches for DM. Thermal freeze-out in a radiation-dominated (RD) universe can yield the correct DM abundance if the annihilation rate takes the nominal value $\langle \sigma_{\rm ann} v \rangle_{\rm f} = 3 \times 10^{-26}$ cm$^3$ s$^{-1}$ (called ``WIMP miracle"). However, this scenario has come under pressure by recent experiments. For example, Fermi-LAT's results from observations of dwarf spheroidal galaxies~\cite{fermi1} and newly discovered Milky Way satellites~\cite{fermi2} have placed upper bounds below on $\langle \sigma_{\rm ann} v \rangle_{\rm f}$ that are below the nominal value for certain final states. Based on these results, a recent analysis~\cite{Beacom} has ruled out thermal DM with a mass below 20 GeV in a model-independent way (unless there is $P$-wave annihilation or co-annihilation). For specific annihilation channels, thermal DM with a mass up to 100 GeV can be excluded.

The situation can change in a non-standard thermal history where the universe is not RD at the time of freeze-out~\cite{KT}. An important example is an epoch of early matter domination (EMD), which is a generic feature of early universe models arising from string theory constructions (for a review, see~\cite{KSW}). In this context, an EMD era is driven by modulus fields that are displaced from the minimum of their potential during inflation and come to dominate the energy density of the post-inflationary universe due to their long lifetime. Moduli eventually decay and form a RD universe prior to big bang nucleosynthesis (BBN). Thermal freeze-out or freeze-in during EMD can accommodate the observed DM relic abundance for $\langle \sigma_{\rm ann} v \rangle_{\rm f} <3 \times 10^{-26}$ cm$^3$ s$^{-1}$~\cite{GKR,Erickcek}. While a small annihilation rate leads to DM overproduction in a RD universe, entropy generation at the end of an EMD phase can regulate the overabundance and bring it down to an acceptable level.   

String constructions involve many modulus fields that can lead to multiple stages of EMD separated by phases of RD. In the standard picture, each period of EMD is driven by a single field with the last one being the most relevant for DM production. However, it is possible that two (or more) fields are simultaneously present during the last epoch of EMD. We study such a ``two-field" scenario and show that the presence of a second field, even if it constitutes a tiny fraction of the energy density and decays very quickly, can significantly enhance the temperature of the universe during EMD. We calculate the abundance of DM particles produced via freeze-out/in under such a modification and find that it opens up the allowed parameter space toward considerably larger DM masses. As a result, PeV-scale DM can be comfortably accommodated by an EMD phase that reheats the universe to a temperature at or below 10 GeV.         

The rest of this paper is organized as follows. In Section II, we briefly review the standard ``single-field" scenario of EMD and its consequences for DM production. In Section III, we discuss the two-field scenario for EMD and its various regimes. In Section IV, we discuss DM production via thermal freeze-out/in in the two-field scenario. In Section V, we present the main results of this paper. We conclude the paper with some discussions in Section VI. Some of the details of our calculations are included in the Appendix.

%%%%%%%%%%%%%%%%%%%%%%%%%%%%%%%%%%%%%%%%%%%%%%%%%%%
\section{Early Matter Domination: the Standard Lore}

An era of EMD can arise from oscillations of a long-lived scalar field $\phi$ with mass $m_\phi$ and decay width $\Gamma_\phi$.\footnote{It is also possible that EMD is driven by non-relativistic quanta produced in the post-inflationary universe~\cite{Ng,Hooper,Scott2}.} Such a field is typically displaced from the true minimum of its potential during inflation. It starts oscillating about the minimum when the Hubble expansion rate is $H_{\rm osc} \simeq m_\phi$. Since these oscillations behave like matter, the ratio of their energy density to that of background radiation increases proportional to the scale factor $a$. 
They will therefore come to dominate the energy density of the universe leading to an epoch of EMD. Assuming that $\phi$ decays perturbatively, which is valid if its couplings to other fields are sufficiently small and its potential is not too steep, its oscillations decay when the Hubble expansion rate is $H_{\rm R} \simeq \Gamma_\phi$ and result in a RD universe with the following reheat temperature:
\be \label{treh}
T_{\rm R} \simeq \left({90 \over \pi^2 g_{*,{\rm R}}}\right)^{1/4} \left(\Gamma_\phi M_{\rm P}\right)^{1/2} . 
\ee
Here $g_{*,{\rm R}}$ is the number of relativistic degrees of freedom at temperature $T_{\rm R}$, and $M_{\rm P}$ is the reduced Planck mass.

The decay of $\phi$ is a continuous process and, assuming that decay products are relativistic and thermalize immediately, it forms a subdominant thermal bath during EMD that grows in time. For $H \gg \Gamma_\phi$, the instantaneous temperature $T$ of the thermal bath follows~\cite{GKR,Erickcek}:
\be \label{tinst}
T \approx \left({6 g^{1/2}_{*,{\rm R}} \over 5 g_*}\right)^{1/4} \left({30 \over \pi^2}\right)^{1/8} \left(H T^2_{\rm R} M_{\rm P}\right)^{1/4} , 
\ee
where $g_*$ denotes the number of relativistic degrees of freedom at temperature $T$. 

For small DM annihilation rates, $\langle \sigma_{\rm ann} v \rangle_{\rm f} < 3 \times 10^{-26}$ cm$^3$ s$^{-1}$, the correct DM abundance can be obtained via thermal freeze-out/freeze-in during EMD~\cite{GKR,Erickcek}.\footnote{Another possibility for obtaining the correct abundance is direct production of DM particles in $\phi$ decay~\cite{KMY,GG,ADS}. This scenario can be embedded in explicit string compactifications~\cite{ACDS1} (see~\cite{ACDS2,ADM} for dark radiation and inflationary constraints on this embedding).} The relic abundance due to freeze-out is given by:  
\be \label{fodens}
\left(\Omega_{\chi} h^2 \right)^{1-{\rm field}}_{\rm f.o.} \simeq 1.6 \times 10^{-4} ~ {g^{1/2}_{*,{\rm R}} \over g_{*,{\rm f}}} ~ \left({m_\chi/T_{\rm f} \over 15}\right)^4 ~ \left({150 \over m_\chi/T_{\rm R}}\right)^3 ~ \left({3 \times 10^{-26} ~ {\rm cm}^3 ~ {\rm s}^{-1} \over \langle \sigma_{\rm ann} v \rangle_{\rm f}}\right) ,
\ee
where $g_{*,{\rm f}}$ is the number of relativistic degrees of freedom at $T = T_{\rm f}$. If $\langle \sigma_{\rm ann} v\rangle_{\rm f}$ is very small, then DM particles will never reach thermal equilibrium at $T > m_\chi$. In this scenario, the DM relic abundance is due to freeze-in of DM production from annihilations of the standard model (SM) particles. The main contribution arises from production at $T \sim m_\chi/4$~\cite{GKR}. DM particles produced at higher temperatures are quickly diluted by the Hubble expansion (when $\langle \sigma_{\rm ann} v \rangle_{\rm f}$ has none or mild dependence on the temperature), while production at lower temperatures is Boltzmann suppressed. The relic abundance due to freeze-in, assuming that $\chi$ represents one degree of freedom, is given by~\cite{GKR}:
\be \label{fidens}
\left(\Omega_{\chi} h^2 \right)^{1-{\rm field}}_{\rm f.i.} \simeq 0.062 ~ {g^{3/2}_{*,{\rm R}} \over g^3_{*}(m_\chi/4)} ~ \left({150 \over m_\chi/T_{\rm R}}\right)^5 ~ \left({T_{\rm R} \over 5 ~ {\rm GeV}}\right)^2 ~ \left({\langle \sigma_{\rm ann} v \rangle_{\rm f} \over 10^{-36} ~ {\rm cm}^3 ~ {\rm s}^{-1}}\right) ,
\ee
where $g_*(m_\chi/4)$ is the number of relativistic degrees of freedom at $T = m_\chi/4$.

For a given DM mass, when the freeze-out and freeze-in abundances become comparable, it signals a transition between the two regimes. The annihilation rate at which the transition occurs can be roughly estimated by setting the expressions in Eq.~(\ref{fodens}) and Eq.~(\ref{fidens}) equal. However, for an accurate calculation of this transition one needs to solve a set of Boltzmann equations that also include details of the thermalization of DM particles (including their kinetic equilibrium) and other species with sizable interactions with DM.

%%%%%%%%%%%%%%%%%%%%%%%%%%%%%%%%%%%%%%%%%%%%%%%%%%%% 
\section{Early Matter Domination: the Two-field Scenario}\label{EMD2}

We now consider a situation where two fields $\phi$ and $\varphi$ with corresponding energy densities $\rho_\phi$ and $\rho_\varphi$ are present during the EMD. We define the parameters $f$ and $\alpha$ as follows:
\be \label{def}
f \equiv {\rho_{\varphi,{\rm i}} \over \rho_{\phi,{\rm i}}} ~ ~ ~ ~ ~ , ~ ~ ~ ~ ~ \alpha \equiv {\Gamma_\varphi \over \Gamma_\phi}.
\ee
Here, $\Gamma_\phi$ and $\Gamma_\varphi$ are the decay widths of $\phi$ and $\varphi$ respectively, and $\rho_{\phi,{\rm i}}$ and $\rho_{\varphi,{\rm i}}$ denote the initial energy density in $\phi$ and $\varphi$ respectively. 

We are interested in a situation where both $\phi$ and $\varphi$ are present during an epoch of EMD as opposed to two separate phases of EMD driven by $\phi$ and $\varphi$ respectively. Therefore, without loss of generality, we consider the case where $f < 1$ and $\alpha > 1$ with $\alpha f \gg 1$. \footnote{The case with $f > 1$ and $\alpha < 1$ leads to a similar scenario with the roles of $\phi$ and $\varphi$ exchanged. On the other hand, the case with $f > 1$ and $\alpha > 1$ results in successive phases of EMD driven by $\varphi$ and $\phi$ respectively, and the case with $f < 1$ and $\alpha < 1$ leads to a similar scenario with $\varphi$ and $\phi$ switching roles.}      

In order to find the instantaneous temperature of the thermal bath, we need to solve the following system of Boltzmann equations:
\begin{eqnarray} \label{2fieldeqs}
&& {\dot \rho}_\phi + 3 H \rho_\phi = - \Gamma_\phi \rho_\phi \, , \\ \nonumber
&& {\dot \rho}_\varphi + 3 H \rho_\varphi = - \Gamma_\varphi \rho_\varphi \, , \\ \nonumber
&& {\dot \rho}_{\rm r} + 4 H \rho_{\rm r} = \Gamma_\phi \rho_\phi + \Gamma_\varphi \rho_\varphi \, .
\end{eqnarray}
In the absence of the second field (i.e., $\rho_\varphi = 0$), the situation is reduced to the standard EMD scenario with a single field $\phi$. The two-field scenario of EMD has three different regimes:
\vskip 2mm
\noindent
{\bf (1) Two-field regime ($H > \Gamma_\varphi$) -} In this regime, both of the $\phi$ and $\varphi$ fields are present. The right-hand side of Eq.~(\ref{2fieldeqs}) is similar to that in the single-field case with an additional factor of $(1 + \alpha f)$. Thus, assuming that both fields decay to relativistic particles in the same sector, the instantaneous temperature of the thermal bath for $H \gg \Gamma_\varphi$ is given by:
\be \label{tinst2}
T \approx \left({6 g^{1/2}_{*,{\rm R}} \over 5 g_*}\right)^{1/4} \left({30 \over \pi^2}\right)^{1/8} (\alpha f)^{1//4} ~ \left(H T^2_{\rm R} M_{\rm P}\right)^{1/4} , 
\ee
The important point is that even though the field $\phi$ dominates the energy density, the decay of the second field $\varphi$ determines the temperature due to its larger decayed fraction since $\alpha f \gg 1$. As a result, $T$ is enhanced in this regime compared to the single-field scenario~(\ref{tinst}) by a factor of $(\alpha f)^{1/4}$. 
\vskip 2mm
\noindent
{\bf (2) Transition regime ($H_{\rm tran} < H \lesssim \Gamma_\varphi$) -} In this regime, $\varphi$ has completely decayed while $\phi$ is still present. Since $\alpha f \gg 1$, the amount of radiation produced by $\varphi$ decay dominates over that continuously produced by $\phi$ decay. For $H \gg H_{\rm tran}$, see Appendix A, the instantaneous temperature of the thermal bath is given by:
\be \label{tinsttran}
T \approx \left({22.5 \over g^{1/3}_{*,{\rm R}} g_*}\right)^{1/4} ~ \alpha^{-1/6} f^{1/4} ~ \left({H^2 M^2_{\rm P} \over T_{\rm R}}\right)^{1/3} .
\ee
We note the different scaling of temperature in the transition regime $T \propto H^{2/3} \propto a^{-1}$, which implies that temperature is simply redshifted due to expansion of the universe. While the field $\varphi$ is absent in this regime, its memory still persists in the form of radiation that its decay produced. As shown in Appendix A, one can estimate $H_{\rm tran}$ to be: 
\be \label{htran1}
H_{\rm tran} \simeq 0.5 \left({\pi^2 g_{*,{\rm R}} \over 90}\right)^{1/2} \alpha^{2/5} f^{-3/5} {T^2_{\rm R} \over M_{\rm P}}.
\ee
\vskip 2mm
\begin{figure}[ht!]
\centering
\includegraphics[width=0.8\linewidth]{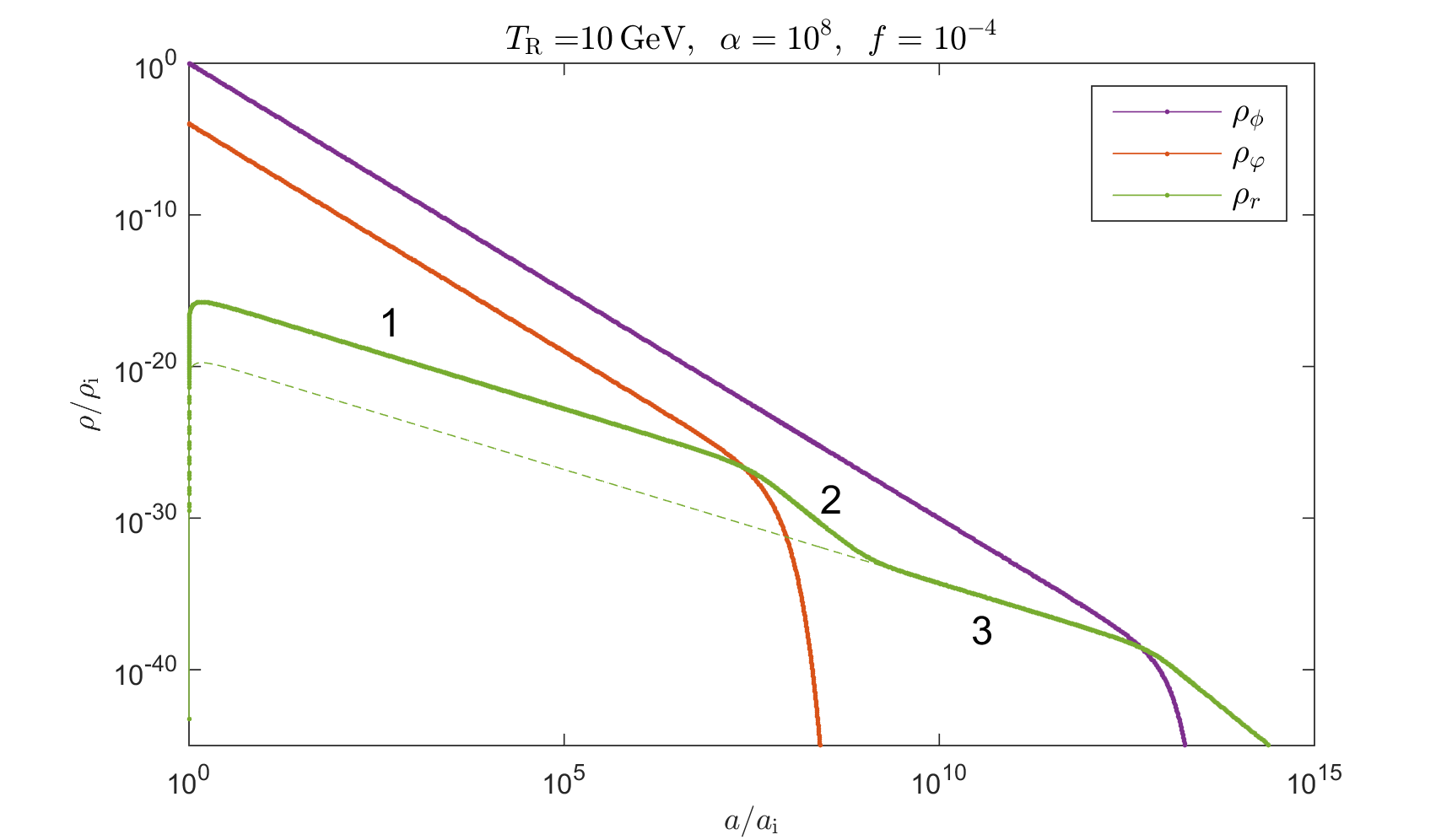}
\captionsetup{justification = raggedright}
\caption{Evolution of energy densities of the dominant field $\phi$ (purple/top), the subdomiant field $\varphi$ (red/middle), and radiation (green/bottom) in a two-field scenario with $f = 10^{-4}$, $\alpha = 10^8$, and $T_{\rm R} = 10$ GeV. Regions 1, 2, and 3 correspond to the two-field, transition, and single-field regimes respectively. The dashed line that extrapolates region 3 denotes the single-field scenario with the same $T_{\rm R}$.}   
\label{fig:rhovsa}
\end{figure}
\noindent
{\bf (3) Single-field regime ($\Gamma_\phi < H \lesssim H_{\rm tran}$ )-} The memory of the second field is erased in this regime and the universe is in the standard EMD phase where temperature is given by the expression in Eq.~(\ref{tinst}).
\vskip 2mm
\begin{figure}[ht!]
  \centering
\includegraphics[width=0.7\linewidth]{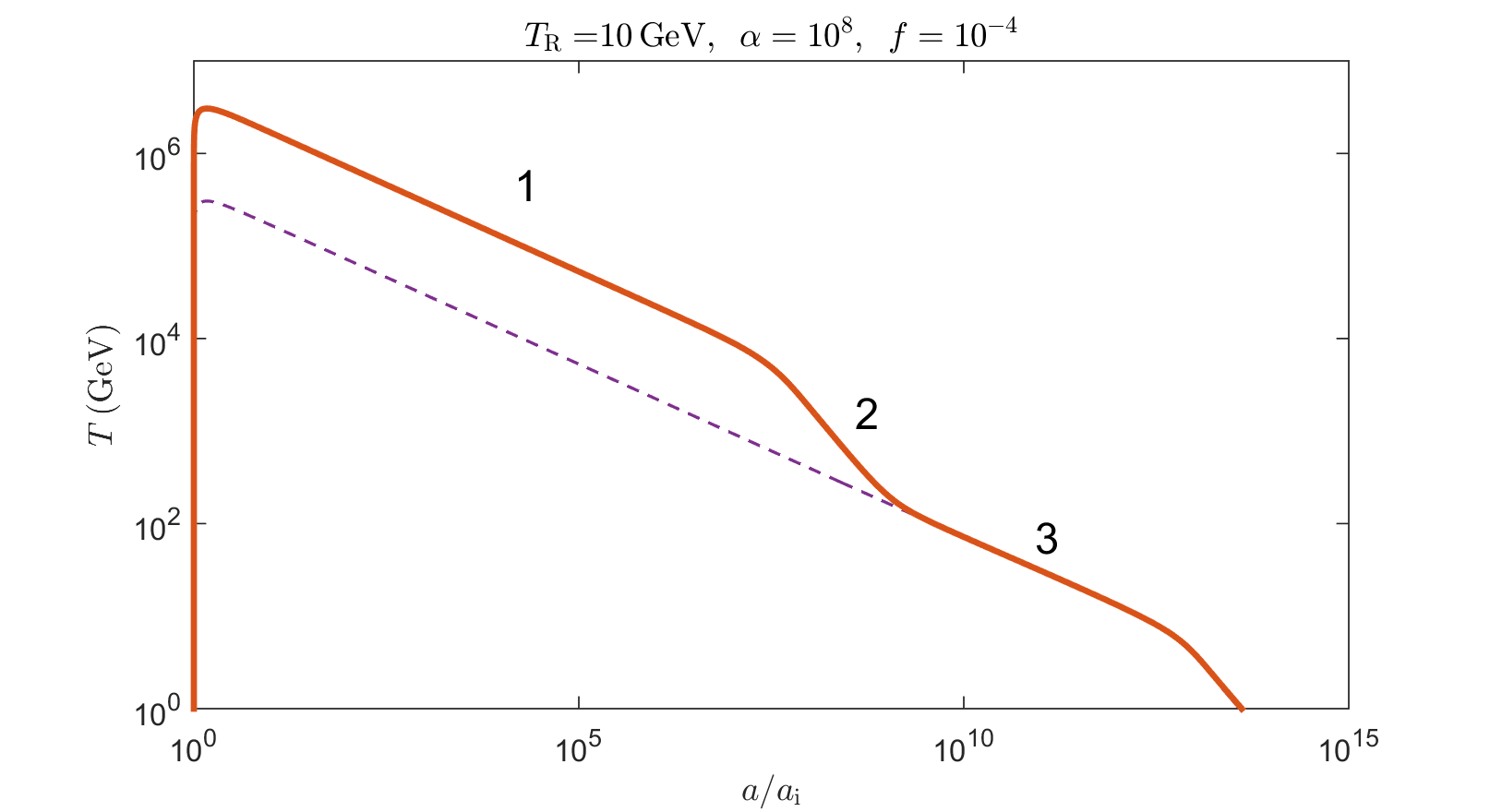}
 \captionsetup{justification = raggedright}
 \caption{Evolution of the temperature in the two-field scenario of Fig.~1. 
%Regions 1, 2, and 3 and the dashed line are the same as in Fig.~1. 
%correspond to the two-field, transition, and single-field regimes respectively. The dashed line that extrapolates region 3 denotes the single-field scenario with %the same $T_{\rm R}$. 
Temperature is enhanced by a factor of $(\alpha f)^{1/4} = 10$ in region 1, approaches that of the single-field scenario in region 2, and coincides with it in region 3.}
  \label{fig:TvsH}
\end{figure}
The important point to note is that the two-field scenario can yield much higher temperatures than that in the single-field scenario as long as $\alpha f \gg 1$. To demonstrate this, we have numerically solved the Boltzmann equations in~(\ref{2fieldeqs}) to find the evolution of the energy densities in the dominant and subdominant fields, $\phi$ and $\varphi$ respectively, and radiation. The initial conditions correspond to the onset of EMD, and hence the initial radiation energy density is negligible. 

In Fig.~1, we show the evolution of the three energy densities (as a function of the scale factor $a$) in a two-field scenario with $f = 10^{-4}$, $\alpha = 10^8$, and $T_{\rm R} = 10$ GeV. We depict the temperature of the universe in this scenario in Fig.~2 and compare it to that in the single-field scenario (i.e., $f = 0$) with the same $T_{\rm R}$. We see that in the two-field regime (region 1), the temperature is enhanced by a factor of $(\alpha f)^{1/4}$. It starts approaching the temperature of the single-field scenario during the transition regime (region 2) as the memory of the second field is being erased. Eventually, this transition becomes complete when the universe enters the single-field regime (region 3).

We would like to reiterate that the enhancement of temperature depends on the product of $\alpha$ and $f$ instead of their individual values. Therefore, as long as $\alpha f \gg 1$, a subdominant field ($f \ll 1$) that decays very early ($\alpha \gg 1$) can indeed significantly enhance the instantaneous temperature at early stages of EMD.

%%%%%%%%%%%%%%%%%%%%%%%%%%%%%%%%%%%%%%%%%%%%%   
\section{Dark Matter Production in the Two-field Scenario}

In this Section, we discuss production of DM via thermal freeze-out/in in the two-field scenario of EMD. We particularly show how the temperature enhancement in the two-field and transition regimes, discussed in the previous Section, affects the DM relic abundance.

In order to calculate the DM relic abundance in the two-field scenario, one needs to solve the equations in~(\ref{2fieldeqs}) together with the following one:
\be \label{dmdens}
{\dot n}_\chi + 3 H n_\chi = - \langle \sigma_{\rm ann} v \rangle_{\rm f}  \left(n^2_\chi - n^2_{\chi,{\rm eq}}\right),
\ee
where $n_{\chi,{\rm eq}}$ denotes the thermal equilibrium value of the DM number density at a given temperature. In the rest of this paper, we consider the case where $\langle \sigma_{\rm ann} v \rangle_{\rm f}$ has no temperature dependence (as happens in the case of $S$-wave dominance). When the annihilation rate is constant, there is no need to have the subscript ``f". We nevertheless keep it for the sake of generality. The situation is qualitatively similar for temperature-dependent $\langle \sigma_{\rm ann} v \rangle_{\rm f}$, but quantitative differences will arise.

In the case of freeze-out, $n_\chi$ closely follows $n_{\chi,{\rm eq}}$ down to the freeze-out temperature $T_{\rm f}$. In the case of freeze-in, we always have $n_\chi \ll n_{\chi,{\rm eq}}$ as DM never reaches thermal equilibrium. The system of four diferentiial equations can be solved numerically in both cases. Here, we provide approximate expressions for the DM abundance in the two-field and transition regimes where the two-field scenarion deviates from the single-field scenario:   
\vskip 2mm
\noindent
{\bf Two-field regime-} Let us first consider freeze-out during the two-field regime. In general, the number density of DM particles at the time of freeze-out follows $n_{\rm f} \propto H_{\rm f}$. The expansion of the universe between freeze-out and reheating, which is the relevant epoch for calculating the entropy density,dilutes $n_{\rm f}$ by a factor of $H^2_{\rm R}/H^2_{\rm f}$. This implies that $\Omega_\chi h^2 \propto H^{-1}_{\rm f}$, which can be seen from Eqs.~(\ref{tinst},\ref{fodens}) in the single-field case. Therefore, after taking into account the additional factor of $(\alpha f)^{1/4}$ in the relation between $T$ and $H$ in the two-field regime~(\ref{tinst2}), we arrive at:
\be \label{fodens2} 
\left(\Omega_\chi h^2 \right)^{2-{\rm field}}_{\rm f.o.} \approx \alpha f \left(\Omega_\chi h^2 \right)^{1-{\rm field}}_{\rm f.o.} ,
\ee
where $\left(\Omega_\chi h^2 \right)^{1-{\rm field}}_{\rm f.o.}$ is given in Eq.~(\ref{fodens}). Due to the same functional dependence of $H$ on $T$, the value of $m_\chi/T_{\rm f}$ is almost the same as that in the single-field case up to a logarithmic term in $\alpha f$. 

Next, we consider freeze-in during the two-field regime. Since $H \propto T^4$, similar to the standard scenario, the bulk of DM particles are produced within one Hubble time when $T_{\rm f} \sim m_\chi/4$. The number density of DM particles at the time of freeze-in is $n_{\rm f} \propto H^{-1}_{\rm f}$ and the dilution factor due to expansion between freeze-in and reheating is $H^2_{\rm R}/H^2_{\rm f}$. This implies that $\Omega_\chi h^2 \propto H^{-3}_{\rm f}$ in this case, which after using Eq.~(\ref{tinst2}) results in:
\be \label{fidens2}
\left(\Omega_\chi h^2 \right)^{2-{\rm field}}_{\rm f.i.} \approx (\alpha f)^3 \left(\Omega_\chi h^2 \right)^{1-{\rm field}}_{\rm f.i.},
\ee
where $\left(\Omega_\chi h^2 \right)^{1-{\rm field}}_{\rm f.i.}$ is given in Eq.~(\ref{fidens}).

We note that the DM relic abundance is enhanced in the two-field regime for both of the freeze-out and freeze-in cases,~(\ref{fodens2}) and~(\ref{fidens2}) respectively, with the latter being more significant. It is then seen from Eqs.~(\ref{fodens},\ref{fidens}) that, for fixed $T_{\rm R}$ and $\langle \sigma_{\rm ann} v \rangle_{\rm f}$, the parameter space that produces the correct DM abundance is shifted toward larger values of $m_\chi$.
%and smaller values of $\langle \sigma_{\rm ann} v \rangle_{\rm f}$ in both the freeze-out and freeze-in cases. 

%
\vskip 2mm
\noindent
{\bf Transition regime-} In the case of freeze-out, the relic abundance of DM particles at reheating follows the usual scaling $\Omega_\chi h^2 \propto H^{-1}_{\rm f}$. However, in the transition regime the relation between $H$ and $T$ is given by the expression in Eq.~(\ref{tinsttran}). After using~(\ref{tinst},\ref{tinsttran}), we find:
\be \label{fodenstran}
\left(\Omega_\chi h^2 \right)^{\rm tran}_{\rm f.o.} \approx  0.15 ~ \left({g_{*,{\rm f}} \over g_{*,{\rm R}}}\right)^{5/8} ~ 
\alpha^{-1/4} f^{3/8} ~ \left({T_{\rm f} \over T_{\rm R}}\right)^{5/2} ~ \left(\Omega_\chi h^2 \right)^{1-{\rm field}}_{\rm f.o.}.
\ee
Due to the different relation between $H$ and $T$, the value of $T_{\rm f}/m_\chi$ differs from that in the two-field regime and the single-field case by logarithmic corrections. 

However, the situation is very different in the case of freeze-in. The comoving number density of DM particles produced via freeze-in is proportional to $\int{n^2_{\chi,{\rm eq}} a^3 dt}$. Starting at a temperature $T \gg m_\chi$, we have $n_{\chi,{\rm eq}} \propto T^3$. In both the two-field regime and the single-field scenario, the $H \propto T^4$ relation, see Eqs.~(\ref{tinst}) and (\ref{tinst2}) respectively, causes the integral to be dominated by the lowest relevant $H$, which corresponds to $T \sim m_\chi/4$~\cite{GKR,Erickcek}.
%\footnote{Production at lower temperatures is negligible because of the Boltzmann suppression of $n_{\chi,{\rm eq}}$ in the non-relativistic regime.} 
On the other hand, in the transition regime, see~(\ref{tinsttran}), we have $H \propto T^{3/2}$. As a result, as shown in Appendix B, the integral is now controlled by the largest value of $H$ in the transition regime, namely $H \simeq \Gamma_\varphi$. Up to an overall proportionality factor, see Appendix B, the freeze-in DM abundance is then found to be:
\be \label{fidenstran}
\left(\Omega_\chi h^2\right)^{\rm tran}_{\rm f.i.} \propto f^{3/2} ~ (T_{\rm R} M_{\rm P}) ~ \left({m_\chi \over 1 ~ {\rm GeV}}\right) ~ \langle \sigma_{\rm ann} v \rangle_{\rm f} .
\ee

An interesting point to note is that the DM abundance in this case has a milder dependence on $m_\chi$ and $T_{\rm R}$ as compared to the two-field regime and the single-field scenario, see Eqs.~(\ref{fidens2},\ref{fidens}). This is because freeze-in production mainly occurs at the onset of the transition regime regardless of the value of $m_\chi$.  

%%%%%%%%%%%%%%%%%%%%%%%%%%%%%%%%%%%%%%%%%%%%%%%
\section{Results}

In this Section, we present our results. We have numerically solved the coupled system of four Boltzmann equations in~(\ref{2fieldeqs},\ref{dmdens}) to obtain the DM relic abundance. The initial conditions are set such that we begin well within the EMD phase, but also long before either of the $\phi$ or $\varphi$ field decays, so that the initial radiation energy density is negligible. This allows us to obtain the behavior due to decay of the two fields, as opposed to the residual effects at the start of EMD. Decayed energy densities are tracked until they are sufficiently small to be unimportant for the subsequent evolution, and are then dropped to facilitate faster numerical calculation. We have taken the detailed temperature dependence of the \(g_*\) factor into account down to $T_{\rm R}$. In order to calculate the DM relic abundance, we have normalized the DM number density with the entropy density long after decay of the dominant field $\phi$. 

We investigate the parameter space, in the $m_\chi-\langle \sigma_{\rm ann} v \rangle_{\rm f}$ plane that yields the correct DM abundance via freeze-out/in for various values of $f$ and $\alpha$, as well as $T_{\rm R}$. Each \(T_{\rm R}\) has a corresponding single-field scenario ($f = 0$) that we use as a baseline for comparison. In Figs.~3 and 4, we show curves in the $m_\chi-\langle \sigma_{\rm ann} v \rangle_{\rm f}$ plane that represent individual choices of the three varied parameters that reproduce the correct abundance. We vary $f$ for constant $\alpha$ in Fig.~3, and $\alpha$ for constant $f$ in Fig.~4. The left and right panels in each figure correspond to $T_{\rm R} = 10$ GeV and $T_{\rm R} = 1$ GeV respectively. For a given set of parameters, DM is underproduced (overproduced) above/outside (under/inside) each curve. The peak of each curve marks the transition between freeze-in (on the left) and freeze-out (on the right).

The curves, in general, consist of three distinct regions that correspond to DM production in regions 1, 2, or 3 of Section \ref{EMD2}. The central region that encompasses the peak of each curve, mimics the shape of the single-field curve while being offset toward higher DM masses and slightly smaller annihilation rates. This distinguishes the part of the parameter space where DM production happens well within the two-field regime (region 1). The curves then move into a near-vertical transition region on both the freeze-in and freeze-out sides, which is identified with DM production in the transition regime (region 2). The two ends of each curve finally merge with the single-field curve, where DM production occurs in the single-field regime after the memory of the second field has been erased (region 3). 
%We label the three regions in Figs. 3 and 4 with the numbers 1, 2, and 3. 
The following main features are observed in the figures: 
\vskip 2mm
\noindent
{\bf (1)} The peak, corresponding to region 1, is more significant for larger values of $f$ and $\alpha$. For fixed values of $f$ and $\alpha$, the shape of the peak does not depend on $T_{\rm R}$, but for higher $T_{\rm R}$ it occurs at a larger $m_\chi$.    
\vskip 2mm
\noindent
{\bf (2)} As $f$ increases for constant \(\alpha\), see Fig.~3, region 1 broadens, pushing out region 2 toward smaller (larger) values of $\langle \sigma_{\rm ann} v \rangle_{\rm f}$ on the freeze-in (freeze-out) side. The change is larger on the freeze-in side.  
\vskip 2mm
\noindent
{\bf (3)} As \(\alpha\) increases for constant \(f\), see Fig.~4, the points where regions 2 and 3 meet are independent of $\alpha$ on the freeze-in side, and only have a mild $\alpha$-dependence on the freeze-out side, moving toward smaller $\langle \sigma_{\rm ann} v \rangle_{\rm f}$. The width of region 1 changes slightly.    
\vskip 2mm

\begin{figure}[ht!]
  \centering
  \subfloat{\includegraphics[width=0.5\textwidth]{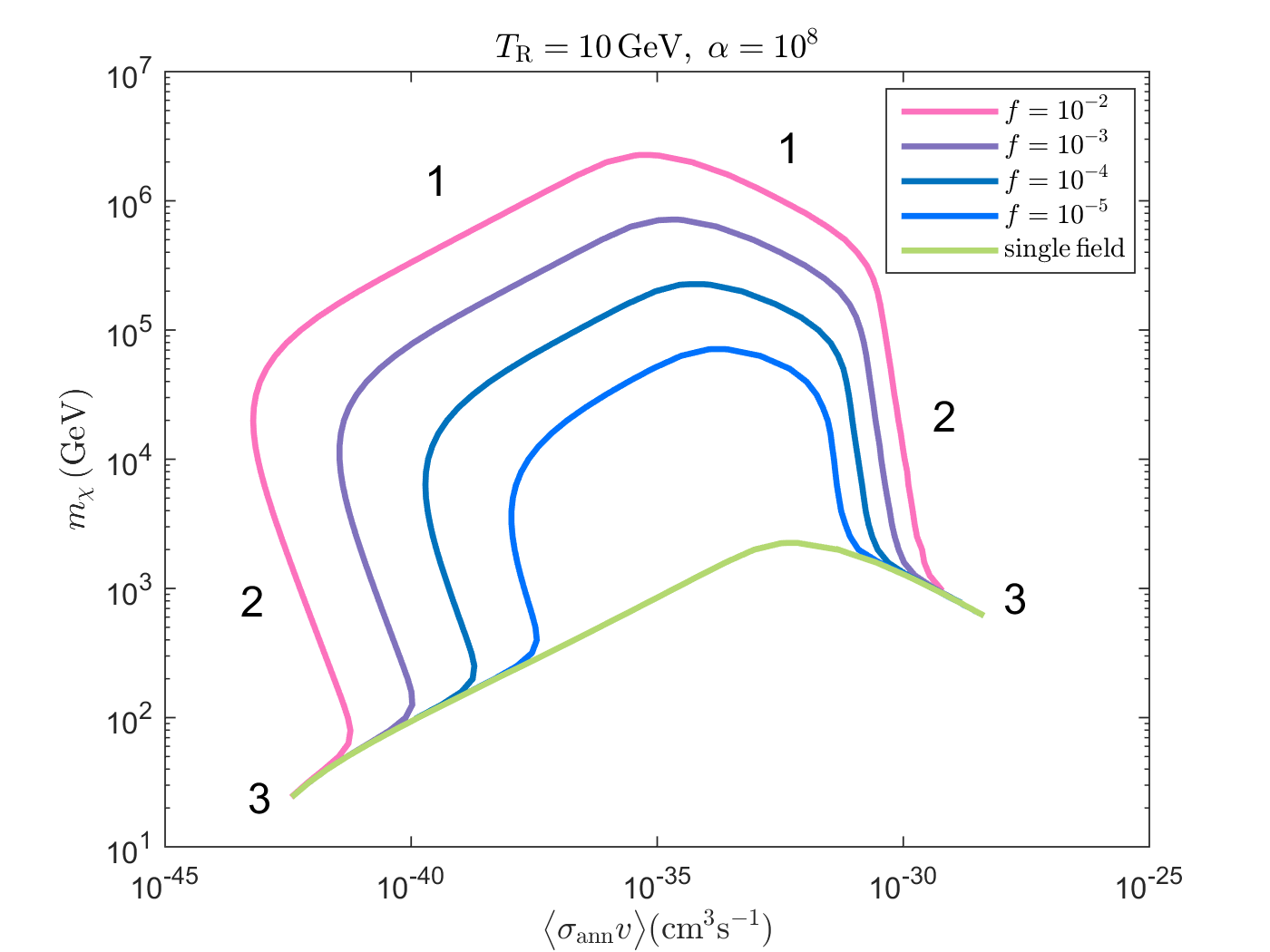}}
  \subfloat{\includegraphics[width=0.5\textwidth]{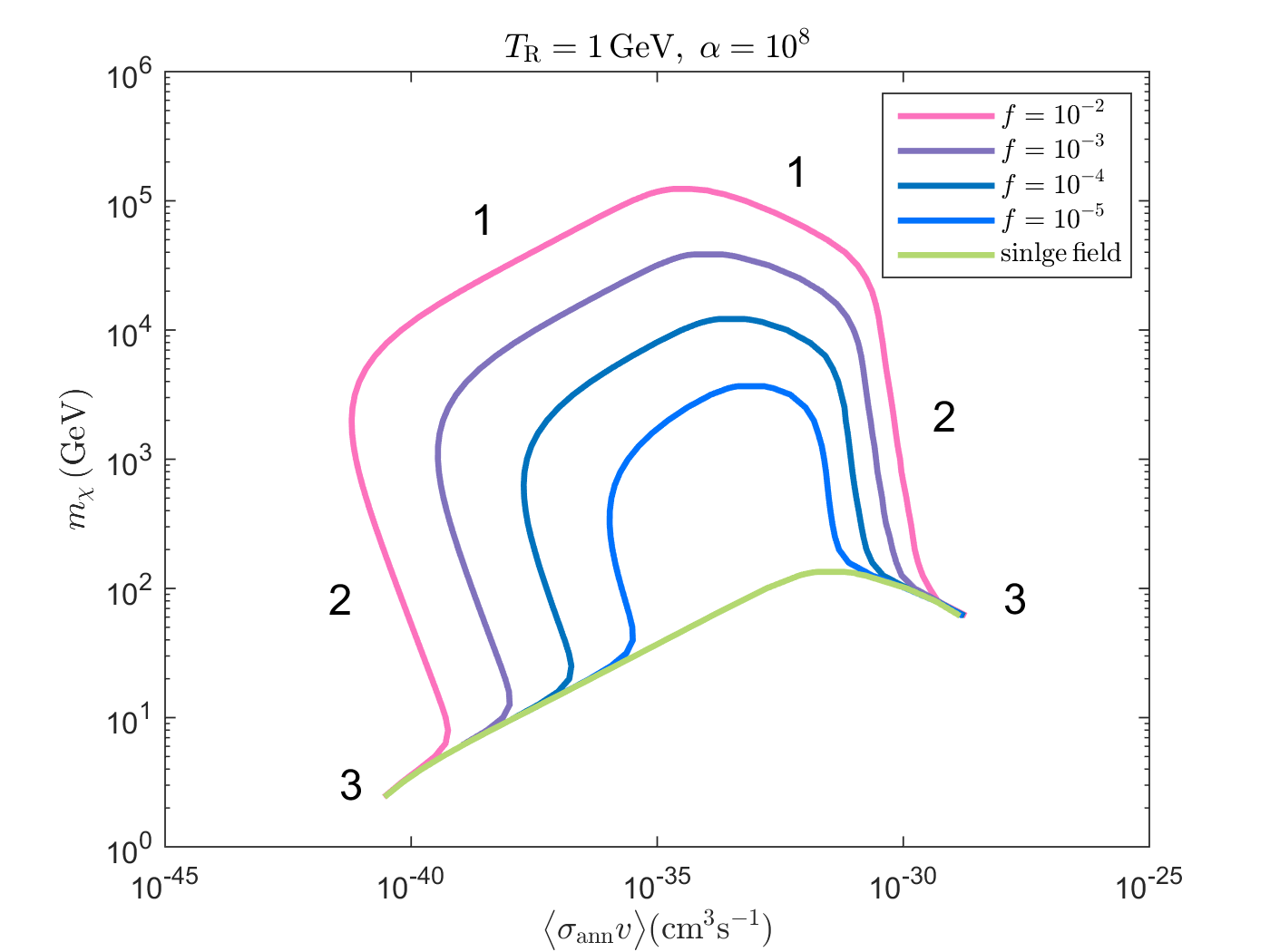}}
 \captionsetup{justification = raggedright}
 \caption{Curves represent points in the $m_\chi-\langle \sigma_{\rm ann} v \rangle_{\rm f}$ plane where the two-field scenario yields the correct DM abundance. We have chosen $\alpha = 10^8$ and varied $f$ between $10^{-2}$ (pink/top) and $f = 10^{-5}$ (blue/bottom) in this figure. The single-field scenario is shown at the very bottom for comparison. The left (right) panel corresponds to $T_{\rm R} = 10$ GeV ($T_{\rm R} = 1$ GeV). DM abundance is set during the two-field regime, transition regime, and single-field regime in regions 1, 2, and 3 respectively. The left and right sides of the curves correspond to freeze-in and freeze-out production respectively.}
\label{fig:mvssigmaf}
\end{figure}
\begin{figure}[ht!]
  \centering
  \subfloat{\includegraphics[width=0.5\textwidth]{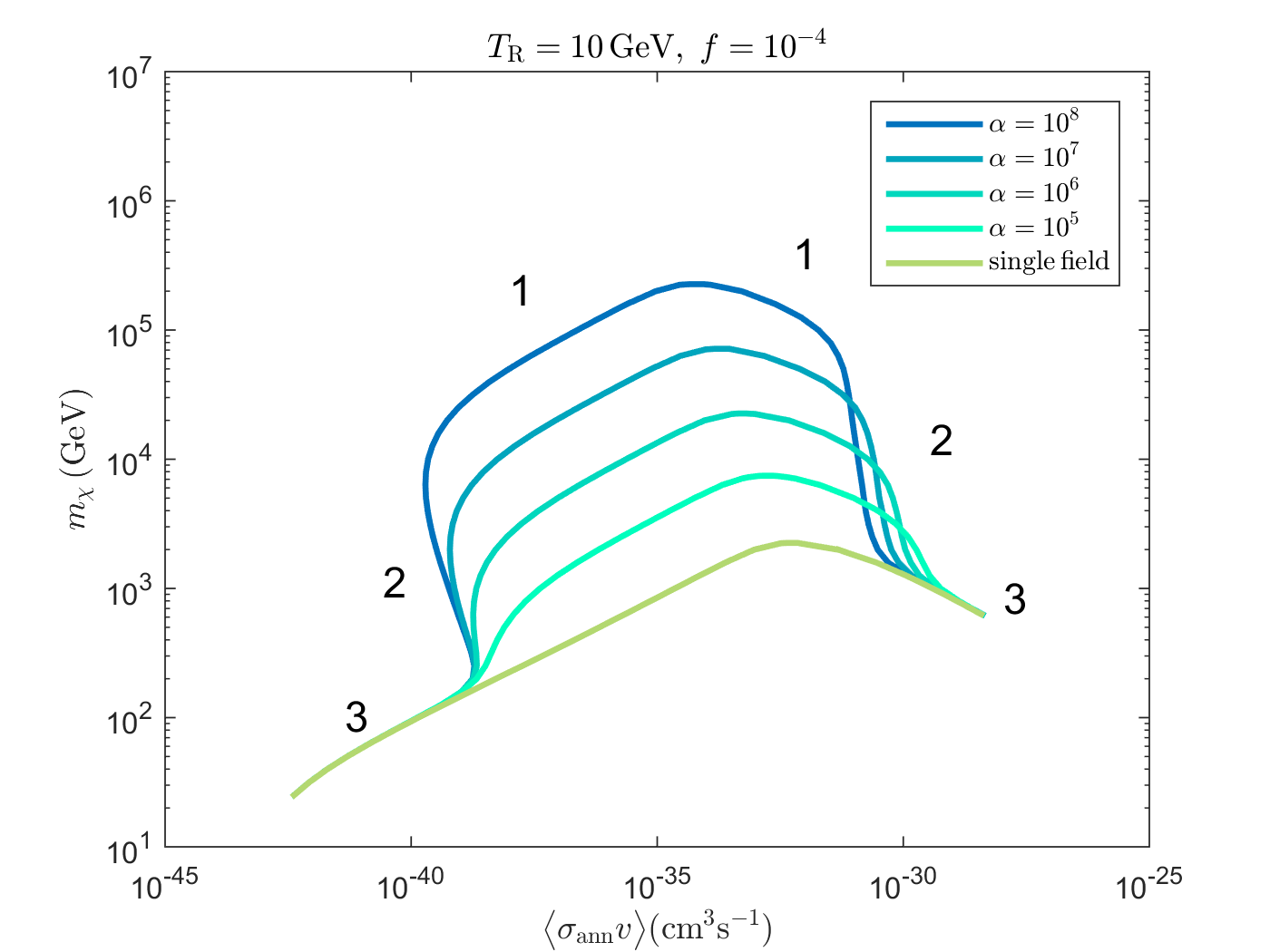}}
  \subfloat{\includegraphics[width=0.5\textwidth]{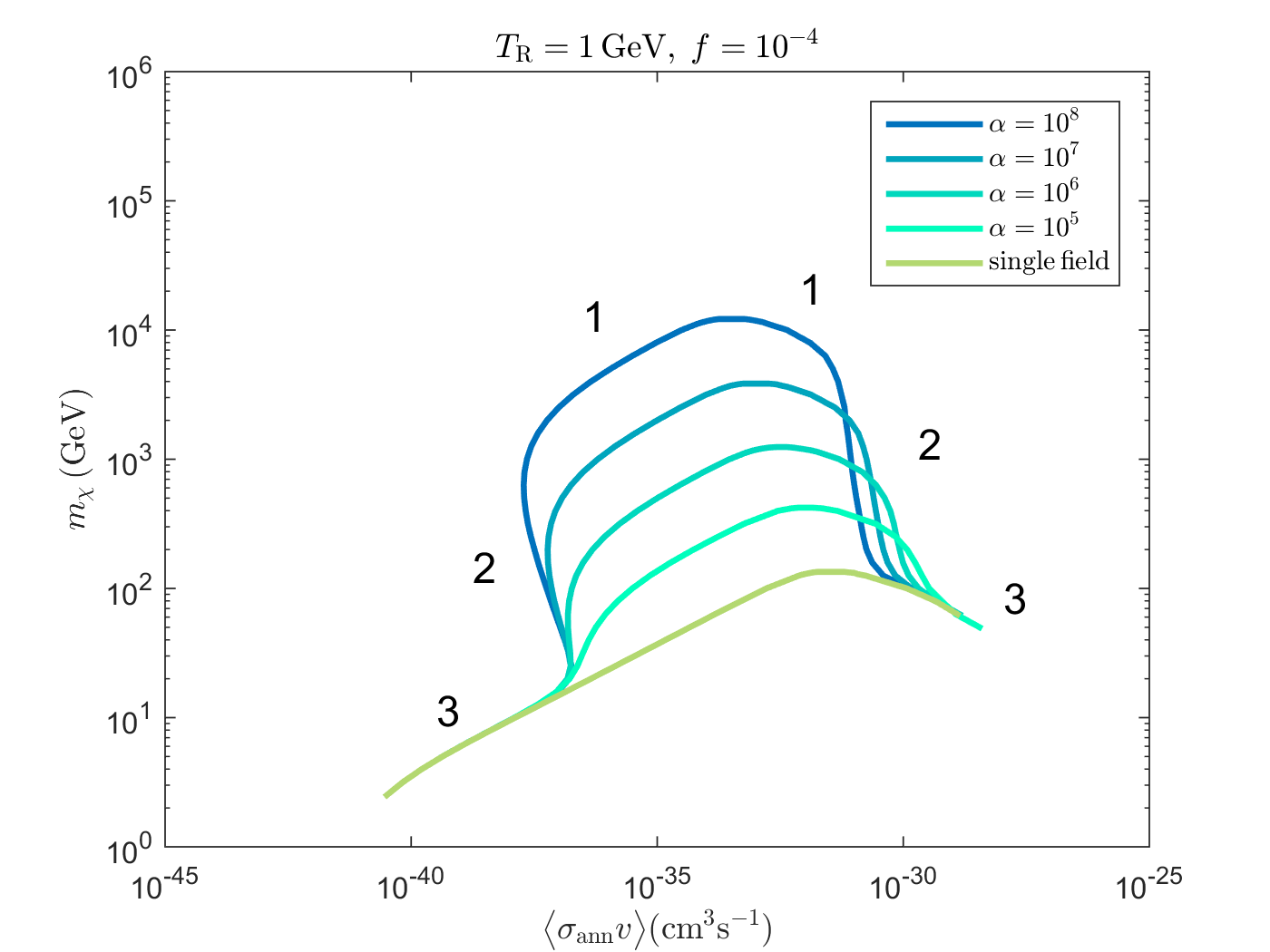}}
  \captionsetup{justification = raggedright}
  \caption{The same as Fig.~3, but we have chosen $f = 10^{-4}$ and varied $\alpha$ between $10^{8}$ (blue/top) and $10^{5}$ (cyan/bottom) in this figure.}
  \label{fig:mvssigmaalpha}
\end{figure}

These features can be qualitatively explained by using the relations that we derived in Section IV. Let us start with point (1) from above. As mentioned earlier, the position of the peak can be estimated by setting the freeze-out and freeze-in DM abundances equal. Using Eqs.~(\ref{fodens2},\ref{fidens2}) for the two-field regime (region 1), we find that $m_\chi \propto (\alpha f)^{1/2}$ and $\langle \sigma_{\rm ann} v \rangle_{\rm f} \propto (\alpha f)^{-1/2}$ at the peak. This explains why the peak moves toward larger values of $m_\chi$ and smaller values of $\langle \sigma_{\rm ann} v \rangle_{\rm f}$ with increasing $f$ or $\alpha$. It also implies that the peak position depends on the product of $\alpha$ and $f$. This is confirmed by comparing the curve with $f = 10^{-5}$ in Fig.~3 to that with $\alpha = 10^7$ in Fig.~4 (both having $\alpha f = 10^3$). As far as dependence on $T_{\rm R}$ is concerned, we note that DM production occurs in the two-field regime when $T_{\rm f}$ is larger than the temperature at $H \simeq \Gamma_\varphi \propto T^{1/2}_{\rm R}$. Since $T_{\rm f} \sim m_\chi/4$ for freeze-in and $T_{\rm f} \propto m_\chi$ (up to logarithmic corrections) for freeze-out, higher $T_{\rm R}$ implies larger values of $m_\chi$ in region 1, hence a higher peak.                          

Regarding points (2) and (3), we need to find the points at which regions 2 and 3 meet. On the freeze-in side, this point can be found by setting the expressions in Eqs.~(\ref{fidens}) and (\ref{fidenstran}) equal. This results in $m_\chi \propto f^{-1/4}$ and $\langle \sigma_{\rm ann} v \rangle_{\rm f} \propto f^{-5/4}$ at the intersection point, which is independent of \(\alpha\). This explains why this point moves down and to the left with increasing $f$ in Fig.~3 but does not move in Fig.~4 (where $f$ is kept constant). On the freeze-out side, the intersection point can be found by setting $T_{\rm f} \propto m_\chi$ (up to logarithmic factors) equal to temperature at $H_{\rm tran}$. After using Eq.~(\ref{tinsttran},\ref{htran1}), this results in $m_\chi \propto \alpha^{1/10} f^{-3/20}$ and $\langle \sigma_{\rm ann} v \rangle_{\rm f} \propto \alpha^{-3/10} f^{9/20}$ at the intersection point. This explains why the point moves slowly in Fig.~3 and very little in Fig.~4. The opposite signs in the exponents of $\alpha$ and $f$ explain why the curves on the freeze-out side of Fig.~4 cross while those of Fig.~3 do not. The points where regions 1 and 2 meet can be found similarly to 2 and 3. We have checked that for these points too, our estimates agree with what is obtained from the figures. Finally, (1) and (2) imply that decreasing $f$ lowers the peak and makes region 1 narrower. This is expected as the curves must be reduced to that for the single-field scenario in the $f \rightarrow 0$ limit.       

The main conclusion from our results is that the two-field scenario can yield the correct abundance for much larger DM masses. As seen in Figs.~3 and 4, the maximum DM mass that it can accommodate is larger by an approximate factor of $(\alpha f)^{1/2}$ than that in the single-field scenario. This holds even for a very small value of $f$ as long as $\alpha$ is sufficiently large so that $\alpha f \gg 1$.\footnote{We note that $\alpha$ is bounded from above in order for the second field not to decay before the onset of EMD. This in turn sets a lower limit on the value of $f$ for which the subdominant field can have a significant effect. However, in realistic situations, this lower limit is typically too small to be relevant.} It is indeed interesting that a subdominant field with a tiny fractional energy density that decays very early can affect DM production in a significant way. As seen in Figs.~3 and 4, the two-field scenario with $T_{\rm R} \gtrsim {\cal O}({\rm GeV})$
%with not too large values of $f$ and $\alpha$ 
can yield the correct abundance for DM masses up to ${\cal O}({\rm PeV})$.

%%%%%%%%%%%%%%%%%%%%%%%%%%%%%%%%%%%%%%%%
\section{Discussions and Conclusion}

So far, we have shown how a second field can enhance the temperature of the universe and thereby affect DM production during EMD. We now briefly discuss some possible realizations of the two-field scenario and reasonable ranges of the $f$ and $\alpha$ parameters that can be expected. 

A natural possibility that can arise in string constructions is that $\phi$ and $\varphi$ are both modulus fields. Such models typically contain many moduli with gravitationally suppressed couplings to matter, implying that $\Gamma_\phi \sim m^3_\phi/M^2_{\rm P}$ and $\Gamma_\varphi \sim m^3_\varphi/M^2_{\rm P}$. Assuming that $\phi$ is the lightest modulus, it drives the last phase of EMD relevant for DM production. Obtaining $T_{\rm R} \sim (1-10)$ GeV then requires that $m_\phi \sim (10^6-10^7)$ GeV. Explicit string constructions exist in the context of KKLT~\cite{KKLT} and large volume~\cite{LVS} flux compactifications where the volume modulus arises as the lightest modulus in the desired mass range~\cite{Nilles, Michele}. The second field $\varphi$ can then be one of the heavier moduli that decays before $\phi$. Generic arguments based on effective field theory estimates~\cite{EFT} or explicit calculations~\cite{Explicit} suggest the amplitude of moduli at the onset  of  their oscillations to be $\gtrsim {\cal O}(0.1~{\rm M_{\rm P}})$. This implies that $H \sim m_\phi$ at the onset of EMD, which requires $m_\varphi < 10^{14}$ GeV in order for $\varphi$ to decay during EMD. For $m_\varphi \lesssim 10^3 m_\phi$, the $\alpha$ parameter is in the range shown in Figs.~3 and 4. Due to the Planckian size of the initial amplitude of both fields, we can have $f \sim {\cal O}(1)$, in which case the effect of the second field will be even more prominent than that shown in the figures.

Another possibility is that the second field $\varphi$ belongs to the visible sector a notable example of which is supersymmetric flat directions. These are directions in the field space of supersymmetric extensions of the SM along which the supersymmetry conserving part of the potential identically vanishes at the renormalizable level~\cite{AD}. These fields are typically displaced from the true minimum of their potential in the early universe. The initial amplitude of their oscillations depends on the level of non-renormalizable operator that lifts flatness~\cite{DRT}, and can be much smaller than $M_{\rm P}$. One can then naturally obtain the small values of $f$ in Figs.~3 and 4 if $\phi$ is a modulus and $\varphi$ is a supersymmetric flat direction. Since $\varphi$ has gauge and Yukawa couplings to other fields in this case, it induces a large mass for them that is proportional to the amplitude of its oscillations. As a result, $\varphi$ decay is kinematically blocked until the induced mass has dropped below $m_\varphi$. For $m_\phi \sim (10^6-10^7)$ GeV (as in the previous case) and $m_\varphi \gtrsim {\cal O}({\rm TeV})$ (so that the scale of supersymmetry breaking in the visible sector is not much higher than TeV), the second field decays during EMD and can lead to values of $\alpha$ that are comparable to or higher than those in Figs.~3 and 4.

In passing, we note a more exotic possibility where the subdominant component is not a field but is composed of primordial black holes (PBH). PBH's with a mass ${\cal O}(10^8~{\rm g})$ evaporate before BBN and could form during a very early bout of matter domination~\cite{PBH}. A situation could then arise where a population of light PBH's in an extended mass range constitute the subdominant component of energy density during EMD.   

In summary, we have studied a modification of EMD that contains two (or, perhaps, more) fields. The presence of a second field may be expected in realistic models and can have important consequences. Even a subdominant field with a tiny fractional energy density that decays much earlier than the dominant field can considerably enhance the temperature of the universe and affect freeze-out/in production of DM during EMD. We have shown that this two-field scenario can open up new regions of the parameter space with much larger DM masses. 
%than in the standard single-field scenario. 
Therefore, the details of the EMD epoch should be taken into account for a careful determination of the DM relic abundance.                       

%Finally, let us comment on possible directions to extend our work. A straightfroward extension involves situations with more than two fields during EMD. One %can also consider a more exotic situation where the subdominant component of energy density is not a field but a primordial black hole (PBH). PBH's a mass %$\lesssim 10^8$ g evaporate before BBN and could form during a very early bout of matter domination~\cite{PBH}. It will be interesting to study the case %where a population of light PBH's in an extended mass range constitute the subdominant component of energy density during EMD.   

%%%%%%%%%%%%%%%%%%%%%%%%%%%%%%%%%%%%%%%%%%
\section*{Acknowledgements}

This work is supported in part by NSF Grant No. PHY-1720174. The idea leading to this work was conceived at the Aspen Center for Physics, which is supported by NSF grant PHY-1607761. We wish to thank Joshua Martin for helpful comments regarding numerical computation, and the Campus Observatory at the University of New Mexico for providing computing time.

%%%%%%%%%%%%%%
\section{Appendix}
\subsection{Temperature during the transition regime}

Here, we first derive an expression for the instantaneous temperature of the universe in the transition regime of the two-field scenario discussed in Section III. The last equation in~(\ref{2fieldeqs}), assuming that $H \gg \Gamma_\phi$, results in: 
\begin{equation}
{d (a^4 \rho_{\rm r}) \over dt} \approx \left(1 + \alpha f e^{ - \Gamma_{\varphi} t}\right) ~ \Gamma_{\phi} \rho_{\phi} a^4 .
\end{equation}
Noting that $\rho_{\phi}a^3 \approx {\rm const}$ in this case, and that $a \propto t^{2/3}$ during EMD, we find:
\begin{equation} \label{16}
{d (a^4 \rho_{\rm r}) \over dt} \approx \left(1 + \alpha f e^{ - \Gamma_{\varphi} t} \right) ~ \Gamma_{\phi} \rho_{\phi,{\rm i}} a_{\rm i}^4 ~ \bigg(\frac{t}{t_{\rm i}}\bigg)^{2/3} , 
\end{equation}
where $t_{\rm i}$ is an initial time that we take to be the onset of EMD. Then $\rho_{\rm r,i} = 0$, and integrating both sides of~(\ref{16}) gives:
\begin{equation}
a^4 \rho_{\rm r}  
%= \Gamma_\phi \rho_{\phi,{\rm i}} a_{\rm i}^4 t_{\rm i}^{-2/3} \int_{t_i}^t t'^{2/3} (1 + \alpha f e^{-\Gamma_{\varphi} t'}) dt' \equiv 
\approx  \Gamma_\phi \rho_{\phi,{\rm i}} a_{\rm i}^4 ~ t_{\rm i}^{-2/3} ~ \left(I_1 + \alpha f I_2 \right),
\end{equation}
where
\begin{equation}
I_1 \equiv \frac{3}{5} (t^{5/3} - t_i^{5/3}) ~ ~ ~ ~ , ~ ~ ~ ~ I_2 \equiv 
%\int_{t_i}^t t^{2/3} e^{-\Gamma_{\varphi} t} dt = 
\Gamma_{\varphi}^{-5/3} \left[ \gamma(5/3,\Gamma_{\varphi} t) - \gamma(5/3, \Gamma_{\varphi} t_{\rm i}) \right].
\end{equation}
Here, $\gamma$ denotes the lower incomplete gamma function. We can now solve for \(\rho_r\) and in turn get the corresponding temperature from \(\rho_r = (\pi^2/30) g_{*} T^4\), making use of \(\rho_{\phi,i} \approx 3 H_i^2 M_{\rm P}^2\) and \(t_i = 2/3H_i\):
\begin{equation} \label{fullT(H)}
T \approx \left({40\Gamma_\phi M_{\rm P}^2 \over \pi^2 g_*}\right)^{1/4} ~ \left(\frac{I_1 + \alpha f I_2}{t^{8/3}}\right)^{1/4}
\end{equation}

Since $t \gg \Gamma^{-1}_\varphi$ in the transition regime, and noting that $t_{\rm i} \ll \Gamma^{-1}_\varphi$, the incomplete gamma functions in $I_2$ approach $\Gamma^{-5/3}_\varphi \Gamma(5/3)$ and $(3/5) t^{5/3}_{\rm i}$ respectively, leading to:
\be \label{Gamma}
I_1 + \alpha f I_2 \approx {3 \over 5} t^{5/3} + \alpha f ~ \Gamma^{-5/3}_\varphi \Gamma(5/3) .
\ee
During the transition regime, the second term on the right-hand side of this expression dominates. After using $\Gamma_\varphi = \alpha \Gamma_\phi$ and Eq.~(\ref{treh}), we find:
\be \label{T(H)tra}
T \approx \bigg(\frac{22.5}{g_{*,{\rm R}}^{1/3}g_{*}}\bigg)^{1/4} ~ \alpha^{-1/6} f^{1/4} ~ \bigg(\frac{H^2 M_{\rm P}^2}{T_{\rm R}}\bigg)^{1/3} . 
\ee

The first term on the right-hand side of Eq.~(\ref{Gamma}) will eventually take over as it increases in time. At that point, the expression in~(\ref{fullT(H)}) is precisely reduced to that in the single-field scenario given in~(\ref{tinst}). Therefore, we can approximately find the time after which the effect of the second field completely disappears by equating the two terms on the RH side of~(\ref{Gamma}). This yields: 
\begin{equation} \label{htran2}
H_{\rm tran} \simeq 0.5 \left(\frac{\pi^2 g_{*,{\rm R}}}{90}\right)^{1/2}  \alpha^{2/5} f^{-3/5} \frac{T_{\rm R}^2}{M_{\rm P}} ,
\end{equation}
where $\Gamma_\phi < H \lesssim H_{\rm tran}$ corresponds to the single-field regime.

\subsection{Freeze-in during the transition regime}

Here, we derive the abundance of DM produced via freeze-in during the transition regime. From Eq.~(\ref{dmdens}), noting that \(n_\chi << n_{\chi,{\rm eq}}\) in the case of freeze-in, we find:  
\begin{equation}
\frac{d(a^3n_\chi)}{dt} \approx a^3\langle\sigma_{\rm ann} v\rangle_{\rm f} ~ n_{\chi,{\rm eq}}^2.
\end{equation}
%. 
After converting $dt$ to $dH$, this equation becomes: 
\begin{equation}
\frac{d(a^3n_\chi)}{dH} \approx  {-2 \Gamma^2_\varphi \over 3H^4} a^3_\varphi \langle\sigma_{\rm ann} v\rangle_{\rm f} ~ n_{\chi,{\rm eq}}^2 .
\end{equation}
Here, we have used $t=2/3H$ and $a^3 = a^3_\varphi (\Gamma_\varphi/H)^2$ during EMD, where $a_\varphi$ is the value of the scale factor at the onset of the transition regime $H \simeq \Gamma_\varphi$. Starting at temperatures $T \gg m_\chi$, and assuming that $\chi$ represents one degree of freeedom, the equilibrium number density is \(n_{\chi,eq} = (\zeta(3)/\pi^2) g_\chi T^3\). We thus have: 
\begin{equation}
\frac{d(a^3n_\chi)}{dH} \approx {-2 \zeta(3)^2 \over 3 \pi^4} {T^6 \over H^4} ~ \langle\sigma_{\rm ann} v\rangle_{\rm f} ~ \Gamma^2_\varphi a^3_\varphi. 
\end{equation}
After using Eq.~(\ref{T(H)tra}), this becomes: 
\begin{equation} \label{int}
\frac{d(a^3n_\chi)}{dH} \approx - \bigg(\frac{22.5}{g_{*,{\rm R}}^{1/3}g_{*}}\bigg)^{3/2}\frac{2\zeta(3)^2}{3 \pi^4} ~ \alpha ^{-1} f^{3/2} ~ {M^4_{\rm P} \over T^2_{\rm R}} ~ \langle\sigma_{\rm ann} v\rangle_{\rm f} ~ \Gamma^2_\varphi a^3_\varphi.
\end{equation}
The integral of the RH side over $H$ is controlled by the largest value of $H$ during the transition regime, namely $\Gamma_\varphi$. After using $a^3 = a^3_\varphi (\Gamma_\varphi/H)^2$ once again, and $\Gamma_\varphi = \alpha \Gamma_\phi$, we find: 
\begin{equation} \label{fi}
n_\chi \approx \bigg(\frac{22.5}{g_{*,{\rm R}}^{1/3}g_{*,\varphi}}\bigg)^{3/2} ~ \frac{2\zeta(3)^2}{3\pi^4} ~ f^{3/2} ~ {\Gamma_\phi M^4_{\rm P} H^2 \over T^2_{\rm R}} ~ \langle\sigma_{\rm ann} v\rangle_{\rm f},
\end{equation}
where $g_{*,\varphi}$ is the number of relativistic degrees of freedom at $H = \Gamma_\varphi$. After normalizing this frozen number density by the entropy density at the end of EMD, and using the expression in Eq.~(\ref{treh}), we arrive at: 
\begin{equation}
{n_{\chi} \over s} \approx {(4 g_{*,\varphi})}^{-3/2} ~ \frac{15 \zeta(3)^2}{\pi^3} ~ f^{3/2} ~ (T_{\rm R} M_{\rm P}) ~ \langle\sigma_{\rm ann} v\rangle_{\rm f}.
\end{equation}
This can be directly used to find \(\Omega_\chi h^2\) (where we have dropped an overall proportionality factor):
\be \label{fitran} 
\left(\Omega_\chi h^2 \right)^{\rm tran}_{\rm f.i.} \propto f^{3/2} ~ (T_{\rm R} M_{\rm P}) ~ \left({m_\chi \over 1 ~ {\rm GeV}} \right) ~ 
\langle \sigma_{\rm ann} v \rangle_{\rm f}.
\ee

We note that Eq.~(\ref{fi}) is obtained by integrating the expression in~(\ref{int}) for a constant $\langle \sigma_{\rm ann}v \rangle_{\rm f}$, which we have considered throughout this paper. In cases where $\langle \sigma_{\rm ann} v \rangle_{\rm f} \propto T^n$, with $n > 0$, freeze-in during the transition regime yields a higher DM abundance. The enhancement is more significant for a strong temperature dependence of $\langle \sigma_{\rm ann} v \rangle_{\rm f}$, like models studied in~\cite{Garcia, Amin}.

%%%%%%%%%%%%%%

\end{document}